\documentclass[twocolumn,showpacs,aps,prl,preprintnumbers,superscriptaddress]{revtex4}
\usepackage{graphicx}
\usepackage{dcolumn}
\usepackage{epsfig}
\usepackage{amsmath}
\RequirePackage{xspace}

\input babarsym.tex

\def\epem  {\ensuremath{e^+e^-}\xspace}

\newcommand{\thetaT}{\ensuremath{\theta_{\rm T}}}
\newcommand{\costhr}{\ensuremath{\cos\thetaT}}

\newcommand{\xf}{\mbox{${\cal F}$}}
\def\KS    {\ensuremath{K^0_{\scriptscriptstyle S}}}
\def\Bu      {\ensuremath{B^+}}
\def\Bub     {\ensuremath{B^-}}
\def\Bbar    {\overline{B}{}}

\def\Bzb     {\ensuremath{\Bbar^0}}
\def\Bz      {\ensuremath{B^0}}
\def\BpBm    {\ensuremath{\Bu  \Bub}}
\def\BzBzb   {\ensuremath{\Bz  \Bzb}}
\newcommand{\DE}{\ensuremath{\Delta E}}
\newcommand{\pvec}{{\bf p}}

 \def\mes{\mbox{$m_{\rm ES}$}}

\newcommand{\aone}       {\mbox{$a_1$}}

\newcommand{\api}     {\mbox{$a_1 \pi$}}
\newcommand{\aoppiz}  {\mbox{$a^+_1 \pi^0$}}

\newcommand{\aozpip}  {\mbox{$a^0_1 \pi^+$}}

\newcommand{\aoneToThreePi}{\mbox{$a_1 \rightarrow 3 \pi$}}

\newcommand{\aoneToRhoPi} {\mbox{$ a_1 \rightarrow \rho \pi  $}}
\newcommand{\aonepToPiPiPi} {\mbox{$ a^+_1 \rightarrow \pi^- \pi^+ \pi^+$}}

\newcommand{\aonezToPiPiPi} {\mbox{$ a^0_1 \rightarrow \pi^- \pi^+ \pi^0$}}

\newcommand{\aonepmMassToThreePip}   {\mbox{$ a^{\pm}_1(1260) \rightarrow (3\pi)^{\pm}  $}}

\newcommand{\aonepmMassToPiPiPi} {\mbox{$ a^{\pm}_1(1260) \rightarrow \pi^- \pi^+ \pi^{\pm}$}}

\newcommand{\aonepmMassToPiPizPiz} {\mbox{$ a^{\pm}_1(1260) \rightarrow \pi^{\pm} \pi^0 \pi^0$}}
\newcommand{\aonezMassToPiPiPi} {\mbox{$ a^0_1(1260) \rightarrow \pi^- \pi^+ \pi^0$}}

\newcommand{\Bpmtoap} {\mbox{$B^{\pm} \rightarrow a^{\pm}_1\pi^{0}$}}
\newcommand{\Bptoap}  {\mbox{$B^+ \rightarrow a^+_1\pi^{0}$}}
\newcommand{\Bpmtoaz} {\mbox{$B^{\pm} \rightarrow a^0_1\pi^{\pm}$}}
\newcommand{\Bptoaz}  {\mbox{$B^+ \rightarrow a^0_1\pi^+$}}

\newcommand{\Bztoap}{\mbox{$B^0 \rightarrow a^{\pm}_1\, \pi^{\mp}  $}}

\newcommand{\BpmtoapMass} {\mbox{$B^{\pm} \rightarrow a^{\pm}_1(1260)\pi^{0}$}}

\newcommand{\BpmtoazMass} {\mbox{$B^{\pm} \rightarrow a^0_1(1260)\pi^{\pm}$}}

\newcommand{\atwo}     {\mbox{$a_2$}}

\newcommand{\Btoatwo}  {\mbox{$B \rightarrow a_2 \pi$}}

\newcommand{\BtoatwoMass}  {\mbox{$B \rightarrow a_2(1320) \pi$}}

\newcommand{\Bptoatwop} {\mbox{$B^{+} \rightarrow a^{+}_2\pi^{0}$}}
\newcommand{\Bptoatwoz} {\mbox{$B^{+} \rightarrow a^{0}_2\pi^{+}$}}

\newcommand{\UfourS}     {\mbox{$\Upsilon(4S)$}}

\newcommand{\BRBpToapToThreePi}  {\mbox{$\cal B(\Bptoap) \times \cal B(\aonepToPiPiPi)$}}
\newcommand{\BRBpToazToThreePi}  {\mbox{$\cal B(\Bptoaz) \times \cal B(\aonezToPiPiPi)$}}

\newcommand{\BRBpmToapMassToThreePi}  {\mbox{$\cal B(\BpmtoapMass) \times \cal B(\aonepmMassToPiPiPi)$}}
\newcommand{\BRBpmToazMassToThreePi}  {\mbox{$\cal B(\BpmtoazMass) \times \cal B(\aonezMassToPiPiPi)$}}

\newcommand{\brBpToapToThreePi}    {\ensuremath{(13.2\pm 2.7\pm 2.1)\times 10^{-6}}}
\newcommand{\brBpToap} {\ensuremath{(26.4\pm 5.4\pm 4.1)\times 10^{-6}}}
\newcommand{\brBpToaz}    {\ensuremath{(20.4\pm 4.7\pm 3.4)\times 10^{-6}}}

\newcommand{\brBpToapToThreePiStatErr}    {\ensuremath{(13.2\pm 2.7)\times 10^{-6}}}

\newcommand{\brBpToazStatErr}    {\ensuremath{(20.4\pm 4.7)\times 10^{-6}}}

\newcommand{\brBztoap} {\mbox{$33.2\pm 3.8 \pm 3.0 $}}

\def\babar{{\em B}{\footnotesize\em A}{\em B}{\footnotesize\em AR}}
\def\BB{\mbox{$B\overline B\ $}}
\def\pep2{PEP-II}

\newcommand\etal{{\it et al.}}

\newcommand{\plBase}   [1]         {Phys.\ Lett.}

\def\figurebox#1#2#3{%
    \def\arg{#3}%
    \ifx\arg\empty
    {\hfill\vbox{\hsize#2\hrule\hbox to #2{\vrule\hfill\vbox to #1{\hsize#2\vfill}\vrule}\hrule}\hfill}%
    \else
    {\hfill\epsfbox{#3}\hfill}%
    \fi}

\begin{document}

\preprint{\babar-PUB-07/034, SLAC-PUB-12699, arXiv:0708.0050}

\title{\large  \bf\boldmath 
Evidence for charged
$B$ meson decays to $a_1^{\pm}(1260)\pi^0$ and $a_1^0(1260)\pi^{\pm}$}

%
\author{B.~Aubert}
\author{M.~Bona}
\author{D.~Boutigny}
\author{Y.~Karyotakis}
\author{J.~P.~Lees}
\author{V.~Poireau}
\author{X.~Prudent}
\author{V.~Tisserand}
\author{A.~Zghiche}
\affiliation{Laboratoire de Physique des Particules, IN2P3/CNRS et Universit\'e de Savoie, F-74941 Annecy-Le-Vieux, France }
\author{J.~Garra~Tico}
\author{E.~Grauges}
\affiliation{Universitat de Barcelona, Facultat de Fisica, Departament ECM, E-08028 Barcelona, Spain }
% ...
\collaboration{The \babar\ Collaboration}
\noaffiliation

\newpage

\begin{abstract}

We present measurements of the branching fractions for the decays
\BpmtoapMass\ and \BpmtoazMass\ from a data sample of $232 \times
10^6$ \BB\ pairs produced in \epem\ annihilation through the \UfourS\
resonance. We measure the branching fraction \BRBpmToapMassToThreePi=\brBpToapToThreePi\ with 
a significance of $4.2 \sigma$, and the branching fraction \BRBpmToazMassToThreePi=\brBpToaz\
with a significance of $3.8 \sigma$, where the first error quoted is
statistical and the second is systematic. 

\end{abstract}


\maketitle

The rare decays of \B mesons to two--body final states with an
$a_1(1260)$ and a $\pipm,\piz,\Kpm$ or $\KS$ are important processes for
testing theoretical factorization model predictions for  branching
fractions, branching fraction ratios and \CP-violation parameters. The
measurements can be combined with assumptions about SU(3) symmetries
to form upper bounds on $\Delta\alpha=\mid\alpha-\alpha_{\rm
  eff}\mid$, where $\alpha$ is the weak interaction phase $\alpha\equiv
\arg\left[-V_{td}^{}V_{tb}^{*}/V_{ud}^{}V_{ub}^{*}\right]$ of the
Unitarity Triangle~\cite{pipipiDalitz} and $\alpha_{\rm eff}$ is the
measured phase. The difference $\Delta\alpha$ is a measurement of the
poorly known strength of the penguin amplitudes in the decay and can
be used to improve our understanding of the \CP-violating mechanism.

The rare decays \BpmtoapMass\ and \BpmtoazMass\ are expected to be
dominated by $b \rightarrow u \bar{u} d$ contributions. The branching
fraction for \Bztoap\ has been measured to be $(\brBztoap) \times
10^{-6}$~\cite{milan} and this agrees well with the calculation of
Bauer, Stech and Wirbel~\cite{Bauer} within the framework of naive
factorization and assuming $\left|V_{ub}/V_{cb}\right|$ = 0.08.  
A more recent analysis using naive factorization and measured form
factors
predicts branching fractions in the range $(5-11)\times 10^{-6}$ and
$(4-9)\times 10^{-6}$ for \Bpmtoap\ and \Bpmtoaz,
respectively~\cite{Laporta}.  
Previous measurements have placed 90\% confidence level upper limits
of $1.7\times 10^{-3}$ and $9\times 10^{-4}$ on the branching
fractions for \Bpmtoap\ and \Bpmtoaz, respectively~\cite{Argus}, and
recently the \babar\ collaboration reported the first measurements of
the {\em CP}-violating asymmetries in the decay
\Bztoap~\cite{milanCPParams}.

We present measurements of the branching fractions for the two 
charmless $B$ meson decays
\Bpmtoap\ and \Bpmtoaz\ where the final state contains one neutral and
three charged pions. The \aoneToThreePi\  decay
proceeds mainly through the intermediate states $(\pi \pi)_{\rho} \pi$
and $(\pi \pi)_{\sigma} \pi$ \cite{PDG2004}.  We do not distinguish
between the dominant P-wave $(\pi \pi)_{\rho}$ and the S-wave $(\pi
\pi)_{\sigma}$ in the channel $\pi^+ \pi^-$.  Possible background
contributions from \BtoatwoMass\ are investigated. Charge conjugate
modes are implied throughout this paper.

The data were collected with the \babar\ detector~\cite{BABARNIM} at
the PEP-II asymmetric $e^+e^-$ collider. An integrated
luminosity of 211~fb$^{-1}$, corresponding to 232 million \BB\ pairs,
was recorded at the $\Upsilon (4S)$ resonance (``on-resonance'') at a
center-of-mass (CM) energy $\sqrt{s}=10.58~\gev$.  An additional
20~fb$^{-1}$ were taken about 40~MeV below this energy
(``off-resonance'') for the study of continuum background in which a
charm or lighter quark pair is produced.

Charged particles are detected and their momenta measured by the
combination of a silicon vertex tracker, consisting of five layers of
double-sided silicon detectors, and a 40-layer central drift chamber,
both operating in the 1.5-T magnetic field of a solenoid.  The
tracking system covers 92\% of the solid angle in the CM frame.
Charged-particle identification (PID) is provided by the average
energy loss (\dedx) in the tracking devices and by an internally
reflecting ring-imaging Cherenkov detector.
 A $K/\pi$ separation of better than four standard deviations
($\sigma$) is achieved for momenta below 3~\gevc, decreasing to 
$2.5\sigma$ at the highest momenta in the $B$ decay final states.

The off-resonance data together with the Monte Carlo 
(MC) simulations of the signal decay modes, continuum,
\BB\ backgrounds and detector response~\cite{geant4} are used to
establish the event selection criteria and reconstruction efficiency.  The MC signal events are 
simulated as $B^+$ decays to \api\ with \aoneToRhoPi.
The \aone\ and \atwo\ line shapes are generated with EvtGen~\cite{evtgen},
where we use mass and width parameters from Refs.~\cite{milan} and~\cite{PDG2004}.

Two photons with a minimum energy of 30~\mev\ (100~\mev\ for \Bptoaz)
and an invariant mass of 
$120 < m_{\gamma\gamma} < 150$~\mevcc are used to reconstruct the $\pi^0$.  
The intermediate dipion states
$(\pi^+\pi^-)$ or $(\pi^+\pi^0)$ are required to have an invariant mass
of $0.46 < m_{\pi\pi} < 1.1$~\gevcc. We impose PID requirements to cleanly
identify the charged pions and to suppress
con\-ta\-mi\-nation from $a_1 K$. 
We require the invariant mass reconstructed for candidate \aonepToPiPiPi\
and \aonezToPiPiPi\ decays to be $0.8 < m_{a_1} < 1.8$~\gevcc.  

A $B$ meson candidate is characterized kinematically by the
energy-substituted mass $\mes = \sqrt{(s/2 + \pvec_0\cdot
\pvec_B)^2/E_0^2 - \pvec_B^2}$ and energy difference $\DE =
E_B^*-\sqrt{s}/2$, where the subscripts $0$ and $B$ refer to the
initial \UfourS\ and to the $B$ candidate in the lab-frame,
respectively, and the asterisk denotes the \UfourS\ frame.  The
resolutions in \mes\ and in \DE\ are about 3.0 \mevcc\ and 20 \mev,
respectively. Candidates are required to have $5.25 \le \mes \le 5.29$
\gevcc and $|\DE|\le0.2$ GeV. To reduce fake $B$ meson candidates we
require a $B$ vertex $\chi^2$ probability $>$ 0.01.  The absolute
value of the cosine of the angle between the direction of the $\pi$
meson from \aoneToRhoPi\ with respect to the flight direction of the
$B$ in the \aone\ meson rest frame is required to be less than 0.85 to
suppress misreconstructed candidates. The distribution of this
variable is flat for signal and peaks near $\pm 1$ for
misreconstructed candidates.

To reject continuum background, we use the angle $\theta_T$ between
the thrust axis of the $B$ candidate's decay products and that of the rest of the
tracks and neutral clusters in the event, calculated in the CM
frame. The distribution of $\cos{\theta_T}$ is sharply peaked near
$\pm1$ for combinations drawn from jetlike $q\bar q$ pairs and is
nearly uniform for the isotropic $B$ meson decays; we require
$|\cos{\theta_T}|<0.65$.  

The decay mode \Btoatwo\ can also give background
contributions.  It is suppressed by using the angular variable
${\mathcal A}$, defined as the cosine of the angle between
the normal to the plane of the $3\pi$ resonance and the flight
direction of the bachelor pion evaluated in the $3\pi$ resonance rest
frame. Since the \aone\ and \atwo\ have spins of 1 and 2, respectively,
the distributions of ${\mathcal A}$ for these two
resonances differ. We require $|\mathcal{A}|< 0.6$, which reduces the
\atwo\ background by more than a factor of two in both decay channels.

After all the above selections, we have on average 1.20 and 
1.56 candidates per event in events where there is at least one candidate, for \Bptoap\
and \Bptoaz, respectively, and we select the $B$ candidate with the
$(\pi\pi)$ mass nearest to the nominal $\rho$ mass~\cite{PDG2004}.
From the
simulation, we find that this algorithm selects the
correct-combination candidate in \Bptoap\ and \Bptoaz\ in 65\% and
55\% of events containing multiple candidates, respectively.

We use an unbinned maximum-likelihood fit using five
variables to extract the background and signal 
yields of \Bptoap\ and \Bptoaz. We describe
the $B$ decay kinematics with the two variables \DE\ and \mes. We also
include the invariant mass of the $3\pi$ system ($m_{a_1}$), the variable ${\mathcal A}$ 
and a Fisher discriminant \xf.  This discriminant
combines four variables: the angles with respect to the beam axis 
of the $B$ momentum and $B$ thrust axis in the CM frame, and the
zeroth and second angular moments of the energy flow around
the $B$ thrust axis~\cite{milan}. 

The extended likelihood function is
\begin{equation}
{\mathcal L} = \frac{1}{N!}\exp{\left(-\sum_{j}n_{j}\right)}
\prod_{i=1}^N\left[\sum_{j}n_{j}{\mathcal P}_{j}(\vec{x}_i;\vec{\alpha}_j)\right]\!,
\end{equation}

\noindent where $n_j$ is the yield of events for hypothesis $j$
(signal, \atwo, \BB\ charmless, \BB\ charm or continuum) and $N$ is
the number of events in the sample. 
The probabilities
${\mathcal P}_{j}$ are products of probability density functions (PDF)
for each of the independent variables $\vec{x}_i = \left\{\mes, \DE,
m_{a_1}, \xf, {\mathcal A} \right\}$ evaluated for each event $i$. The
$\vec{\alpha}_j$ are the parameters of the distributions in
$\vec{x}_i$. By minimizing the quantity $-\ln{\mathcal L}$ in two
separate fits, we determine the yields for \Bptoap\ and \Bptoaz.

To take into account the relatively large number of misreconstructed signal events, 
the signal is separated into two components, representing the correctly
reconstructed (true) and the self cross-feed (SCF) candidates, with proportions fixed in the fit for each mode.
SCF occurs when a track from an \aoppiz\ or \aozpip\ is
exchanged with a track from the rest of the event. The fraction of
SCF, determined from MC, is 35\% and 44\% for \Bptoap\ and \Bptoaz,
respectively.

In addition to the \atwo, there are three main categories of
backgrounds: \BB\ charmless, \BB\ charm and continuum.  \BB\
backgrounds are studied using MC simulations of \BzBzb\ and \BpBm\
decays, using a large sample equivalent to $\sim0.8$\,ab$^{-1}$.  
There are 17 \BB\ charmless decays for \Bptoap\ and 20 for \Bptoaz\ that contribute as background.  
Those decays with similar distributions are grouped to form 13 and 10 hypotheses,
respectively, 
and are included in the fit with a fixed
yield as determined from MC.  The total \BB\ charmless yields are $368\pm92$ and
$755\pm164$ for \Bptoap\ and \Bptoaz, respectively.  These are dominated by 
$B\rightarrow\rho\rho$, $B\rightarrow\aone\rho$ 
and the other $B\rightarrow\aone\pi$ mode under study.
The \BB\ charm backgrounds are included as a single hypothesis, with
the normalization of the \BB\ charm yield as a free parameter.
Continuum events come from light quark production.  We establish the
functional forms and parameter values of the PDFs for \BB\ charm and
\BB\ charmless backgrounds from MC simulations.  
For continuum, we use off-resonance data for the Fisher, on-resonance
data with $|\DE|>0.1\gev$ for \mes, and on-resonance data with $5.25 < \mes
< 5.27\gevcc$ for the other variables.

We model the distributions using appropriate functions. The 
${\mathcal A}$ distributions are modeled with polynomials.
For the true signal component, the remaining distributions are
fitted using modified Gaussians
~\cite{crystalBall}, and a relativistic Breit-Wigner line-shape
with a mass-dependent width~\cite{WA76}, as necessary. The SCF component
and the
\atwo\ have similar shapes to the true signal but have broader or more
asymmetric distributions and shifted means.
The \BB\ backgrounds and continuum distributions are modeled with modified Gaussians, 
polynomials, nonparametric functions~\cite{Keys} and, for \mes, 
a phase-space-motivated empirical function~\cite{argus}. 
The PDF
variables are assumed to be independent except for \Bptoaz, where a
two dimensional nonparametric PDF~\cite{Keys} in $m_{a_1}$ and \DE\
accounts for observed correlations in the MC for both
true signal events and SCF.

In the fit there are six free parameters: four yields (signal,
continuum, $a_2$ and \BB\ charm background), and two continuum
background parameters (\DE\ polynomial coefficient and \mes\ shape
coefficient $\xi$ \cite{argus} ).

For \Bptoap, there are $24608$ events in the data sample. We measure 
the raw signal yield to be $459 \pm 78$ events
with a reconstruction efficiency of $12.5\pm0.1\%$, corrected for
differences in tracking and neutral particle reconstruction between
data and MC.  The yield of the decay \Bptoatwop\ is $28 \pm 65$ events.
For \Bptoaz, there are $33375$ events in the data sample and  
we measure the raw signal yield to be $382 \pm 79$ events with a 
corrected reconstruction efficiency of $7.2\pm0.1\%$.
The yield of the decay \Bptoatwoz\ is $107 \pm 65$ events. 

We confirm our fitting procedure by generating and
fitting MC samples containing signal and background populations using the yields as found from
data. We identify a signal yield bias for \Bptoap\ and
\Bptoaz\ of $16.8\pm0.1\%$ and $10.9\pm0.1\%$, respectively. We fit for the branching fractions 
taking into account the fitted signal yield,
the yield bias, the corrected reconstruction efficiency, daughter
branching fractions, and the number of produced $B$ mesons, assuming
equal production rates of \BzBzb\ and \BpBm\ pairs. The statistical
significance is taken as the square root of the difference between the
value of $-2\ln{\cal{L}}$ for zero signal and the value at its minimum. We
measure the branching fraction \BRBpToapToThreePi\ = \brBpToapToThreePiStatErr\ with a
statistical significance of $5.3\sigma$ and the branching fraction
\BRBpToazToThreePi\ = \brBpToazStatErr\ with a statistical
significance of $4.7\sigma$, where the errors are statistical.

\begin{figure}[!htb]
\resizebox{\columnwidth}{!}
{ \includegraphics[]{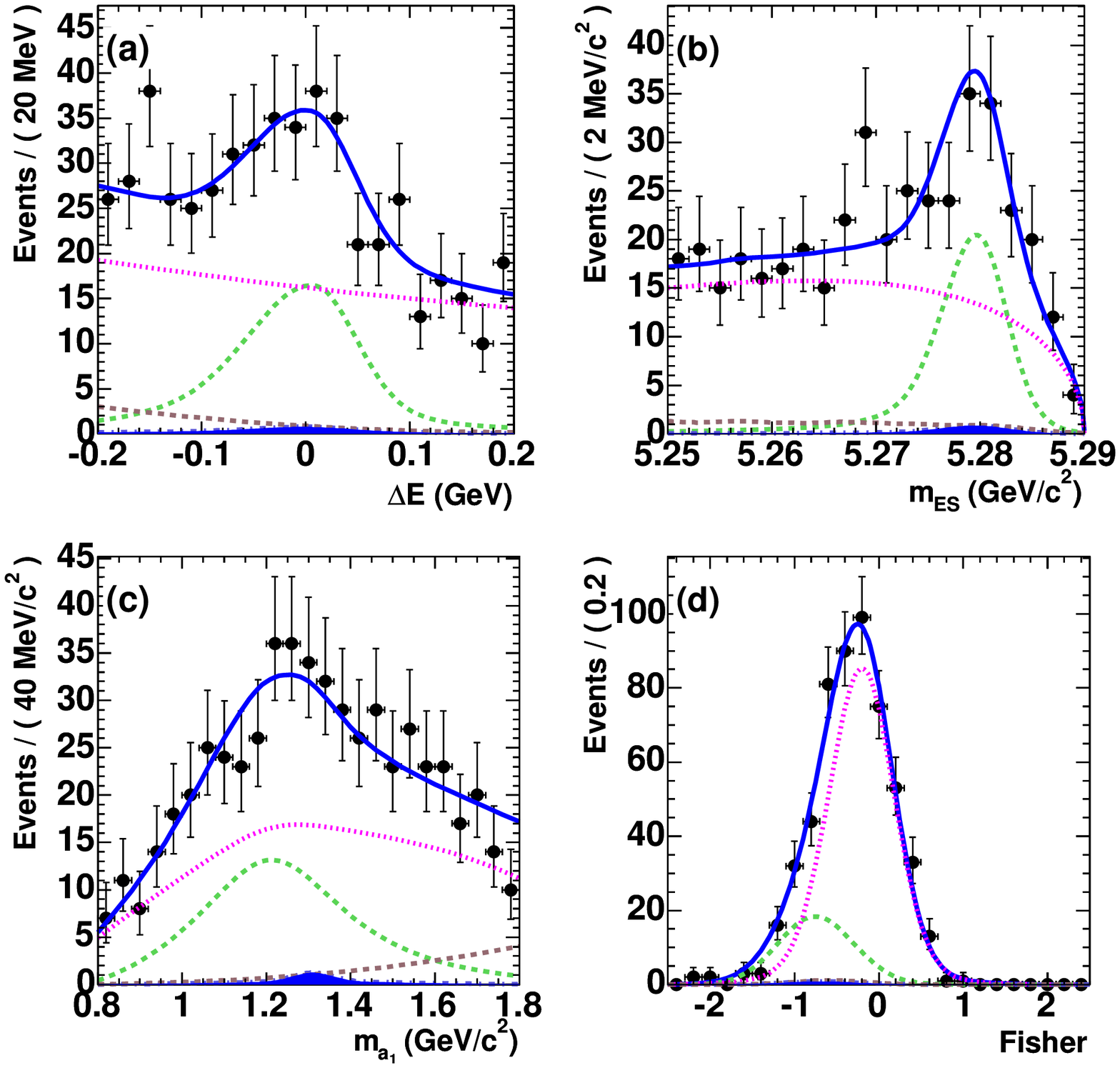} }
\caption{(color online). Projections of a) \DE, b)  \mes, c) $m_{a_1}$, and d) \xf\ 
for \Bptoap. 
Points represent on-resonance data, dashed lines the signal, dotted lines 
the continuum, dashed-dotted lines the \BB\ charm background, the filled region the \atwo\ background
and solid lines the full fit function. These plots are made with a 
requirement on the signal likelihood to enhance the signal, and thus do not 
show all events in the data sample.}
\label{fig:projections1}
\end{figure}

\begin{figure}[!htb]
\resizebox{\columnwidth}{!}
{ \includegraphics[]{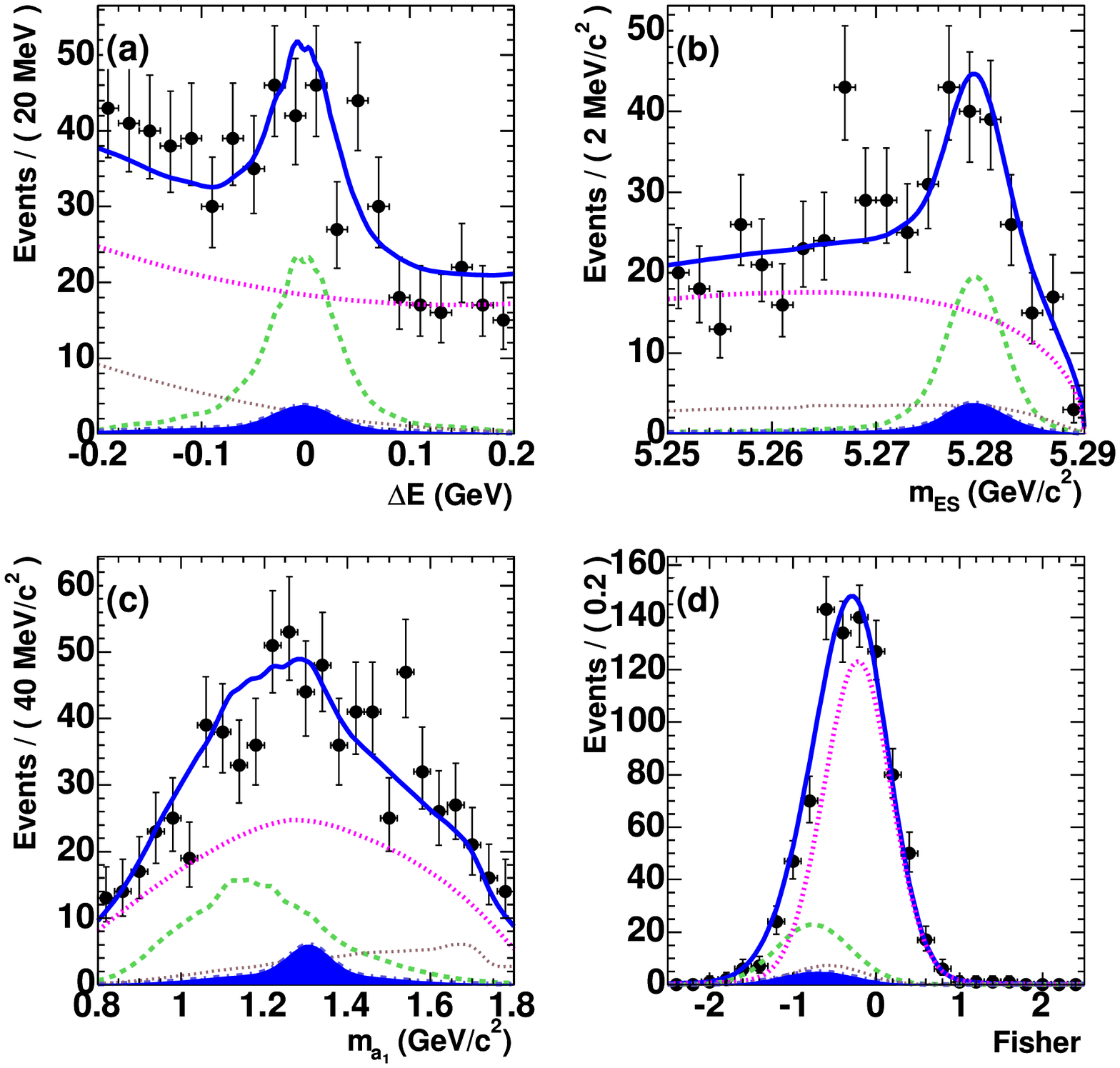}  }
\caption{(color online). Projections of a) \DE, b) \mes, c) $m_{a_1}$, and d) \xf\
  for \Bptoaz, using the same criteria and line styles as Fig.~\ref{fig:projections1}.}
\label{fig:projections2}
\end{figure}

Figs.\ \ref{fig:projections1} and \ref{fig:projections2} show the \DE,
\mes, $m_{a_1}$, and \xf\ projections for \Bptoap\ and \Bptoaz\ made
by selecting events with a signal likelihood (computed without the
variable shown in the figure) exceeding a threshold that optimizes the
expected sensitivity.

\begin{table}[!htb]
\caption{Summary of systematic errors for the \aoppiz\ and \aozpip\ branching fraction measurements.}
\label{tab:systematics}
\begin{center}
\begin{tabular}{lcc}
\hline
Systematic                         & \aoppiz                 & \aozpip  \\ 
\hline
PDF Parameter Variation            & 8.6\%                   & 8.8\%    \\
Fit Bias                           & 8.4\%                   & 5.5\%    \\
$a_1-a_2$ Interference             & 6.6\%                   & 7.4\%    \\       
SCF Variation                      & 4.4\%                   & 8.2\%    \\
Tracking Efficiency                & 3.9\% 		     & 3.9\%    \\
$\pi^0$ Efficiency                 & 3.0\%                   & 3.0\%    \\ 
Flight Direction Criteria          & 2.0\%                   & 2.0\%    \\ 
P-wave and S-wave Reconstruction   & 1.6\%                   &  -       \\
Charmless \BB\ Background          & 1.4\%                   & 3.1\%    \\
Number of \BB\ Pairs               & 1.1\%                   & 1.1\%    \\
$\cos\theta_T$ Selection Criteria  & 1.1\%                   & 1.8\%    \\
Track Multiplicity                 & 1.0\%                   & 1.0\%    \\
$\rho\pi\pi$, $4\pi$ Cross-Feed    & 0.9\%                   & 0.5\%    \\ 
$a_1 K$ Cross-Feed                 &   -                     & 0.4\%    \\
\hline
Total                              & 16\%                  & 16\%       \\ 
\hline
\hline
\end{tabular}
\end{center}
\end{table}

The systematic errors are summarized in Table~\ref{tab:systematics}.
We determine the sensitivity to the
parameters of the signal and background PDF components by varying 
these within their uncertainties.  
The effect of varying the mass and width of the \aone\ 
by the errors as reported in Ref.~\cite{milan} is 
included in the PDF parameters' variation systematic.
The uncertainty in the fit bias correction is taken as half of the fit
bias correction.  The effect of possible interference between \atwo\
and \aone\ is estimated by adding the \atwo\ and \aone\ amplitudes
together with a varying phase difference and using half the maximum change in
yield as an uncertainty.  
The uncertainty in SCF is investigated by varying the SCF fraction.
We also perform a separate fit treating the SCF as an independent background component. 
The fitted branching fraction is compatible with the nominal fit within the increased statistical uncertainty, 
but the statistical significance is reduced to $3.5\sigma$ and $3.0\sigma$ for \Bptoap\ and \Bptoaz,
respectively. 
A systematic uncertainty of
$1.6\%$ is estimated for the difference in reconstruction efficiency
in the decay modes through the dominant P-wave $(\pi\pi)_{\rho}$ and
the S-wave $(\pi\pi)_{\sigma}$.  An error is assigned for the
uncertainty in the fixed charmless \BB\ background yields and possible
interference effects by varying the individual
components by the reported error on the branching fractions~\cite{PDG2004}.  The systematic
errors for the flight direction criteria, number of \BB pairs, \costhr\ selection criteria, track
multiplicity, potential backgrounds from $\rho\pi\pi$ and $4\pi$, and
$a_1 K$ cross-feed are small.
The total systematic error for both modes is $16\%$. 
The significance of the branching fractions, combining
both statistical and systematic errors, is $4.2\sigma$ for \Bptoap\, and $3.8\sigma$ for
\Bptoaz.

In conclusion, we have measured the branching fractions $\cal B(\BpmtoapMass) \times \cal B(\aonepmMassToPiPiPi)$
= \brBpToapToThreePi\ and $\cal B(\BpmtoazMass) \times \cal B(\aonezMassToPiPiPi)$ =
\brBpToaz. Neglecting isoscalar contributions to the two-pion state,
we assume $\cal B(\aonepmMassToPiPiPi)$
is equal to $\cal B(\aonepmMassToPiPizPiz)$ and $\cal B(\aonepmMassToThreePip)$ is
equal to 100\%~\cite{PDG2004}, resulting in $\cal B(\BpmtoapMass)$ =
\brBpToap.
We measure $\cal B(\BpmtoazMass)$ = \brBpToaz, 
assuming $\cal B(\aonezMassToPiPiPi)$ is equal to 100\%.
The first errors quoted
are statistical and the second are systematic. 
The signals are seen with significances 
of $4.2 \sigma$ and $3.8 \sigma$, respectively,  and are in agreement
with factorization model predictions~\cite{Bauer}.

We are grateful for the excellent luminosity and machine conditions
provided by our \pep2\ colleagues, 
and for the substantial dedicated effort from
the computing organizations that support \babar.
The collaborating institutions wish to thank 
SLAC for its support and kind hospitality. 
This work is supported by
DOE
and NSF (USA),
NSERC (Canada),
CEA and
CNRS-IN2P3
(France),
BMBF and DFG
(Germany),
INFN (Italy),
FOM (The Netherlands),
NFR (Norway),
MIST (Russia),
MEC (Spain), and
STFC (United Kingdom). 
Individuals have received support from the
Marie Curie EIF (European Union) and
the A.~P.~Sloan Foundation.

\end{document}